\documentclass{article}
\usepackage{graphicx}
\usepackage{blindtext}
\usepackage{amsmath}
\usepackage{mathtools}
\usepackage{multirow}
\usepackage{pdflscape}
\usepackage{ragged2e}
\usepackage{lineno}
\usepackage[hidelinks]{hyperref}
\usepackage{bm}
\usepackage{upgreek}
\usepackage{bbold}
\usepackage{url}

\DeclareMathAlphabet{\mathpzc}{OT1}{pzc}{m}{it}

\makeatletter
\let\saved@includegraphics\includegraphics
\AtBeginDocument{\let\includegraphics\saved@includegraphics}
\makeatother

\newcommand\icarus{{Icarus}}
\newcommand\ssr{{Space Science Reviews}}
\newcommand\planss{{Planetary and Space Science}}

\begin{document}

\begin{center}
\textbf{\Large{Mapping Saturn using deep learning}}  \\

\noindent I. P. Waldmann$^{1,*}$ \& C. A. Griffith$^{2}$

\noindent {\small{ \textit{$^{1}$Department of Physics \& Astronomy, University College London, Gower Street, WC1E6BT London, United Kingdom \\
$^{2}$University of Arizona, Lunar and Planetary Laboratory, 1629 E University Blvd, Tucson, AZ 85721-0092 \\
$^*$Corresponding author}}} \\

Nature Astronomy, DOI: 10.1038/s41550-019-0753-8
\end{center}

\justify

\textbf{
Clouds and aerosols provide a unique insight into the chemical and physical processes of gas-giant planets. Mapping and characterising the spectral features indicative of the cloud structure and composition, enables an understanding of a planet's energy budget, chemistry and atmospheric dynamics (e.g. \cite{2016Icar..277..196F,2011Sci...332.1413F,2016Icar..271..400B,2016NatCo...713262S}). 
Current space missions to solar-system planets produce high-quality data sets, yet the sheer amount of data obtained often prohibits detailed `by hand' analyses. Current techniques mainly rely on two approaches: 1) identify the existence of spectral features by dividing fluxes of two or more spectral channels;  2) perform full radiative transfer calculations for individual spectra. The first method suffers from accuracy whilst the second from scalability to the whole planetary surface.
Here we developed a deep learning algorithm, PlanetNet, able to quickly and accurately map spatial/spectral features across large, heterogeneous areas of a planet. We demonstrate PlanetNet on Saturn's 2008 storm\cite{baines09}, enhancing the scope of the area previously studied. Our spectral-component maps indicate compositional and cloud variations of the vast region affected by the storm showing regions of vertical upwelling, and diminished clouds at the centre of compact storms. This analysis quickly and accurately delineates the major components of Saturn's storm, thereby indicating regions that can be probed deeper with radiative transfer models. }


The Visual and Infrared Mapping Spectrometer (VIMS) on the Cassini space-craft \cite{2004SSRv..115..111B} is a two-channel mapping spectrometer, which obtains spatially resolved spectra across a 64 $\times$ 64 pixel array. In this work we use the near-infrared channel, which spans 256 contiguously sampled band-passes ranging from 0.85 to 5.1 \,$\mu$m with constant   $\delta \lambda$= 0.016 $\mu$m. Hence each data cube (or hyperspectral image) contains 4096 individual spectra. 
The observations of Saturn's 2008 storm\cite{baines09} are particularly well suited for this work as they encompass multiple, adjacent storms, providing a complex atmospheric feature space to be analysed by PlanetNet. In particular, the data cube V1581233933 contains a rare ammonia ice feature, detected by \cite{2004SSRv..115..111B}, which projects a ``S'' shaped feature on Saturn's disk. This data cube, along with two spatially adjacent cubes, has previously been analysed by the spectral band devision method\cite{baines09}, which allows comparison to our approach. In this study we re-analyse the three original cubes along with three additional adjacent data cubes, all of which were obtained on February 9$^{th}$ 2008. 

PlanetNet is capable of non-parametrically identifying faint features in hyperspectral images and once trained on a given feature, able to search for it across highly heterogeneous data sets. The  algorithm consists of two parts: 1) a spectral clustering algorithm to identify an initial feature set; 2) a double-stream deep convolutional neural network (CNN). Here we provide a brief overview of the algorithm. Additional details are given in the Method section. 
Spectral clustering\cite{1238361} is a versatile clustering algorithm suited for high-dimensional vector spaces.
Unlike with more traditional clustering algorithms such as k-nearest neighbours or k-means, spectral clustering is not dependent on the convexity of the individual cluster sets and can identify highly non-convex clusters. We compute the number of clusters in the data and assign a cluster label to each spectrum, as discussed in more detail in the Methods section. 
Once we obtain our initial clusters, we trim all data within one pixel of a cluster edge to avoid edge uncertainties and proceed to train the neural network (NN) using the remaining spectra. 
The NN contains two branches, a spatial and a spectral channel. The spectral branch takes each remaining spectrum and trains a two layer CNN using ReLU (rectified linear unit) activation functions and two pooling layers. Surrounding each spectrum, we compute a 20x20 spatial image by averaging the spectral cube along the spectral axis. This is the input for the spatial channel which otherwise follows the same NN architecture as the spectral channel. By analysing both spectral and spatial information, we can take into account the morphological and spectral signatures of atmospheric features on Saturn. In other words, a dark storm, for example, will have a distinct spectrum and spatial morphology that correlate together. By including these spatial-spectral correlations, the neural network will take all possible information available into account. Finally, the output of both spectra and spatial channels is fed into a fully connected layer which links to the cluster labels via a logistic regression layer. Figure 1 shows a schematic of the full network. Similar network designs have been successfully used in aerial image classification of commercial land-usage (e.g. \cite{rs9030298,7730324}) for a recent review we refer the reader to \cite{8113128}. 

We verify the classification accuracies using two methods: 1) During training, 30\% of the TC are reserved as `test data'. The test data is randomly chosen from the training set and is used to verify the classification accuracy on unseen data during training. Classification on the test data achieves $\sim 90\%$ accuracy. After training, we further test the trained PlanetNet on a resampled version of the TC by rotating the spatial data and interpolating spectral data within each cluster (see Methods for more detail). Here we achieve an accuracy of 93\%, consistent with accuracies obtained from the test data. 



We apply PlanetNet to the data cube V1581233933, and hereafter refer to this data set as the training cube (TC) in which we identify 5 clearly distinguishable clusters of spectral/spatial features. 
We confirm the existence of these clusters by performing a principal component analysis (PCA\cite{Jolliffe}). Whereas PlanetNet clusters spatial/spectral data by their similarity (low statistical distance), PCA decomposes the spectral cube by its variance.
Should distinct spectral regions exist, we expect to see signatures of these clusters also in the variance of the data. PlanetNet clusters are clearly visible in the principal components and we refer the reader to the Method section for a more in-depth discussion. 

The left of Figure\,\ref{fig:specs} presents the spatial extent of the TC and a map of the spectra features, while the right hand side shows the spectral characteristics of the 5 spectral clusters identified. Here, the blue region corresponds to a large stormy region (SR) surrounding the central dark storm (purple/green) and label\,1 denotes the centre of the ``S'' feature.
Each cluster is distinguished by its absorption and scattering characteristics, indicative of the cloud structure and gas composition. Most salient are the spectral differences between the region surrounding the dark storm features (blue region here-forth referred to as SR) in contrast to the unperturbed regions (red/orange), and the unique signatures of the black storms (purple/green). Spectra 1 \& 4 in Figure\,\ref{fig:specs} are examples of spectra belong to the SR region. We find that regions unaffected by the storm (e.g. spectra 2 and 3) display the brightest albedo at 1 - 2$\,\mu$m (Fig.\,\ref{fig:specs}). At these wavelengths, the bands of the well-mixed CH$_4$ are modulated by clouds, the brightness of which indicates high aerosols. In contrast, the region surrounding the storm features (blue) is dimmer at 1 - 2$\,\mu$m, suggesting either lower clouds, or, more likely, given the great extent of the SR region, these are spectrally darker clouds, as postulated by \cite{baines09}. This interpretation indicates that the blue regions encapsulate current and prior storms darkened by the upwelling material of lower albedo.

These SR features, based on their 1 - 2$\,\mu$m spectra and low 5$\,\mu$m flux, contain relatively large particles. The brightest of these features (spectrum 1 in Figure\,\ref{fig:specs}), forms an ``S'' feature, which coincides with the position of the electric discharges measured on February 9, 2008 by Cassini's Radio and Plasma Wave Science (RPWS) instrument\cite{baines09,2005Sci...307.1255G,2007Icar..190..528F,2008SSRv..137..271F}. We find the ``S'' feature to be the proverbial ``tip of the iceberg'' of a much larger region of upwelling. In Supplementary Figures 1 \& 2, we map the L$^2$ distance (i.e. spectral difference, see equation~\ref{equ:d} in Method) between the peak emission of the ``S'' shaped feature and the remaining SR region. We find that across the SR region the ``S'' shaped feature contains the highest 1 - 2\,$\mu$m  but lowest 5\,$\mu$m flux suggesting the particle size to be highest there.   

Baines et al.\cite{baines09} hypothesise that the particles found in the ``S'' feature are condensed NH$_3$ as indicated by colour maps of the continuum (0.93$\,\mu$m), methane (0.90$\,\mu$m) and the NH$_3$ ice feature (2.73$\,\mu$m). Our analysis defines the relative spectrum of the ``S'' and find that the ``S'' spectrum displays an absorption feature at roughly 2.74$\,\mu$m -- 2.85$\,\mu$m. This is part of a broader continuum that resembles the NH$_3$ ice spectrum characteristic of that observed in Jupiter's storms\cite{2002Icar..159...74B}. Similarly, the averaged spectrum of the full SR region indicates a similar although weak ammonia ice feature. 

Saturn's thermal flux, measured long-ward of $\sim$4.5$\,\mu$m also differentiates the ambient from the stormy atmospheres. The SR flux at 5$\,\mu$m exhibits a comparatively low brightness indicative of higher atmospheric opacity at the 5$\,\mu$m wavelength due to particles roughly of sizes 5$\,\mu$m and larger. In particular, the PlanetNet algorithm identifies clusters pertaining to Saturn's dark storm. The associated spectra (5 and 6) indicate notable spectral structure. Both spectra have 1 - 2$\,\mu$m albedos darker than that of the other regions. Relative to the mean storm region (SR), spectra of the dark storms display absorption at 4.5 - 4.9$\,\mu$m characteristic of PH$_3$ and consistent with the visibility of this feature as indicated in radiative transfer (RT) models\cite{2013Icar..226..402S}. PH$_3$ is a condensible species, the abundance of which increases by vertical upwelling at the observable levels.  The enhancement of this species indicates that the storm experiences strong vertical updrafts. Saturn's emission, longward of 4.5$\,\mu$m, differentiates the centre of the storm (Spectrum 5) from its  immediate surroundings (Spectrum 6). The storm centre exhibits a strong flux, indicating the clearest view into the hotter and deep atmosphere, the clarity of which suggests the eye of a storm. In contrast, the 5$\,\mu$m flux surrounding the storm is commensurate with that of the other clusters. 

Supplementary Fig. 1 shows the identified storm region in more detail and map the statistical distance (equation (\ref{equ:d}) in Method) between the peak emission of the ``S'' shape feature (marked as 1) and all other spectra in the SR labelled regions. Similar to Figure~\ref{fig:specs} we plot individual spectra scattered across the SR region and show the spectral differences to the ``S'' feature by plotting their differences. Whereas most spectral bands are invariant, we do find for all spectra that the 1 - 2\,$\mu$m albedo is suppressed and the thermal emission increased compared to the ``S'' shape feature. This furthermore corroborates the conclusion that the ``S'' shaped feature is indeed the peak of a larger upwelling. 
In Supplementary Fig. 3 we show the same statistical distance map as in Supplementary Fig. 1 but taking the centre of the dark storm (point 5 in Figure\,\ref{fig:specs}) as the reference point. This figure hence shows the likeness of spectra compared to the central dark storm signature. The SR region stands out as being distinct from the dark storm signatures, a feature that is further confirmed by the principal component analysis in Supplementary Figures 4 - 6 discussed in the Method section.



Once trained on a hyperspectral cube, PlanetNet allows us to quickly and accurately map salient spectral regions over multiple heterogeneous data sets, spanning a large area of the planet. Whereas our analysis focused on the training cube so far, we now use PlanetNet to map a much larger region by including 5 additional data cubes that encompass the original storm, as well as two smaller storms eastwards. As shown in Figures\,\ref{fig:stich}\,\&\,\ref{fig:stich2} we detect the presence of the SR feature, significantly beyond the spatial coverage initially reported by \cite{baines09}. In addition, we find similar spatial/spectral signatures of the stormy region around a smaller dark storm to the east (42$^{\circ}$W), indicating that areas of upwelling are common around dark storms on Saturn. We provide individual maps of each mapped component in the Supplementary material (Supplementary Figures 7\,-\,10) as well as the statistical distance map of the SR region across all five data sets in Supplementary Fig. 2.



Past and current planetary missions produce a wealth of data, too abundant to be analysed by ``hand''. More traditional data analysis techniques force us to consider only small volumes of data and a global understanding of spatial distributions of spectroscopic features (e.g. clouds on gas giants) is often lost. 
Maps produced by PlanetNet can give us insight into large-scale dynamics of a planet, while identifying regions of interest for more traditional radiative transfer calculations. This technique is significantly more sensitive and robust than simple spectral band subtraction or division and can reveal previously unseen dynamics in the atmospheres of giant planets.
The ability to identify features in data sets markedly different to the training data (both in pixel scale and observed angle) allows this technique to be easily scalable to large, planet-wide mapping of spectral features. PlanetNet can easily be adapted to other data sets and missions, making it a potentially invaluable tool in the global analysis of planetary mission data in the future.  


\noindent All correspondences should be addressed to I.P.W (ingo@star.ucl.ac.uk).

\paragraph*{Acknowledgements} The authors would like to thank the editor and the referees for helping to greatly improve the clarity of this paper.

This project has received funding from the European Research Council (ERC) under the European Union's Horizon 2020 research and innovation programme (grant agreement No 758892, ExoAI) and under the European Union's Seventh Framework Programme (FP7/2007-2013)/ ERC grant agreement numbers 617119 (ExoLights). IPW furthermore acknowledges funding by the Science and Technology Funding Council (STFC) grants: ST/K502406/1 and ST/P000282/1 and support from Microsoft Azure for Research cloud computing. CAG is funded by the University of Arizona.

\paragraph*{Author Contribution} I.P.W developed the PlanetNet algorithm and conducted the data analysis. C.A.G. provided the calibrated Cassini/VIMS data and lead the spectral interpretation of the data. Both authors equally contributed to writing the text.

\paragraph*{Data Availability} The data analysed in this work are available 
through the Planetary Data System (\url{https://pds.nasa.gov}).
PlanetNet is publicly available through the UCL-Exoplanets GitHub page 
(\url{https://github.com/ucl-exoplanets/}). 
In addition, both code and training data used are permanently archived and can be accessed with the permanent link: \url{https://osf.io/htgrn} or the Digital Object Identifier: DOI 10.17605/OSF.IO/HTGRN.



\begin{figure}
\centering
\includegraphics[width=0.8\textwidth]{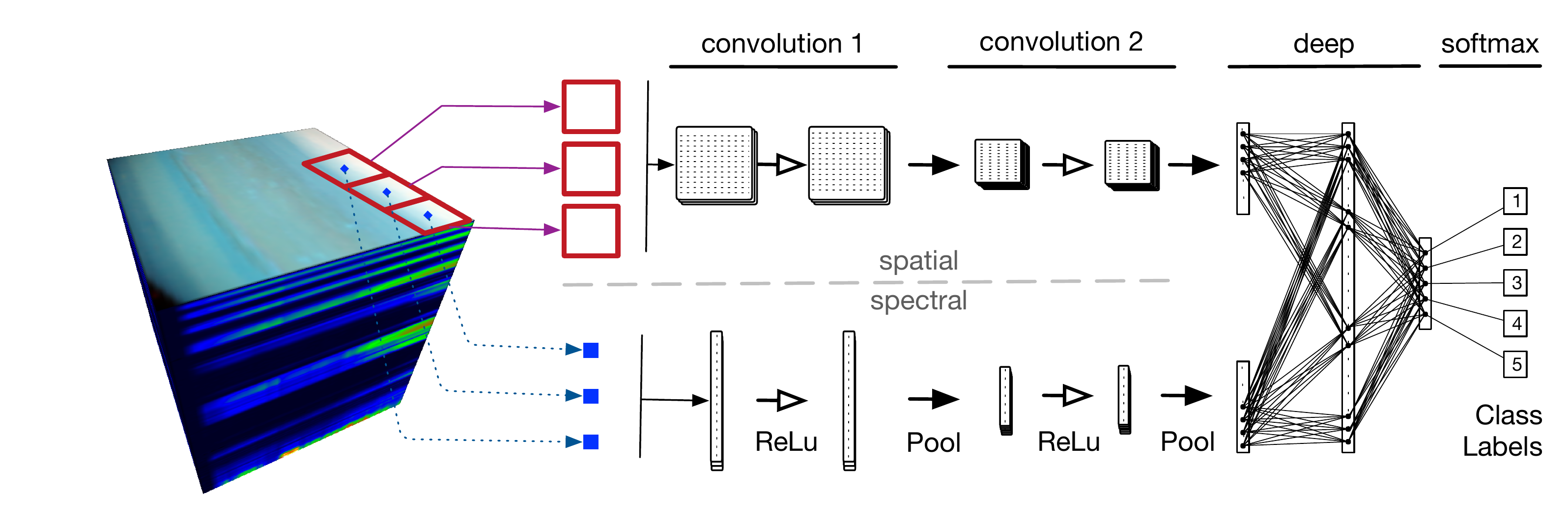}
\caption{{\bf Flowchart of the PlanetNet algorithm}. The blue dots indicate the central pixel at which the spectrum is extracted from the Cassini/VIMS data cube. The red squares indicate the corresponding spatial patches centred on the central spectral pixels. The spatial and spectral data is fed into two convolutional neural networks for the spatial (top network) and spectral (bottom network) information respectively. Both convolutional networks are linked to a fully connected layer combining spatial and spectral convolutional outputs. The output of the fully connected layer is mapped to the class labels. Please see the Method section for more information.  \label{fig:chart}}
\end{figure}

\begin{figure}
\centering
\includegraphics[width=0.8\textwidth]{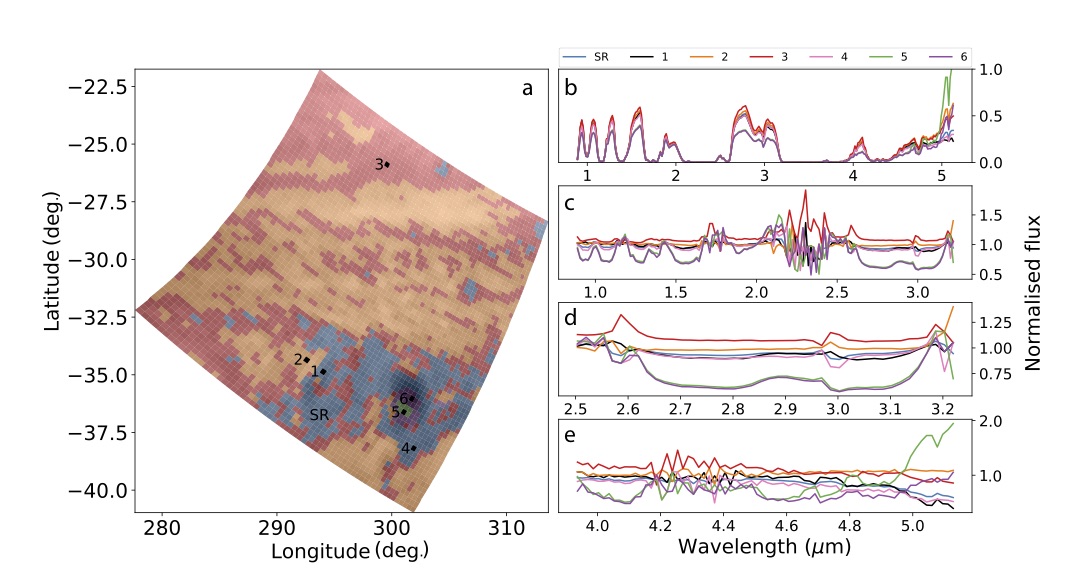}
\caption{{\bf Spatial and spectral characteristics of the PlanetNet identified features.} A: Map of Cassini/VIMS cube V1581233933, coloured according to the 5 different clusters described in the main text:  SR (blue) , 2 (orange), 3 (red), 4 (green), 5 (purple).  B: Atmospheric spectra at locations marked on the map (A), spectrum 1 corresponds to the  ``S'' feature of NH$_3$ ice clouds. The SR spectrum and label correspond to the mean spectrum of the blue designated area on the map. C \& D: Same as B above but with the map's mean spectrum subtracted. \label{fig:specs}}
\end{figure}

\begin{figure}
\centering
\includegraphics[width=\textwidth]{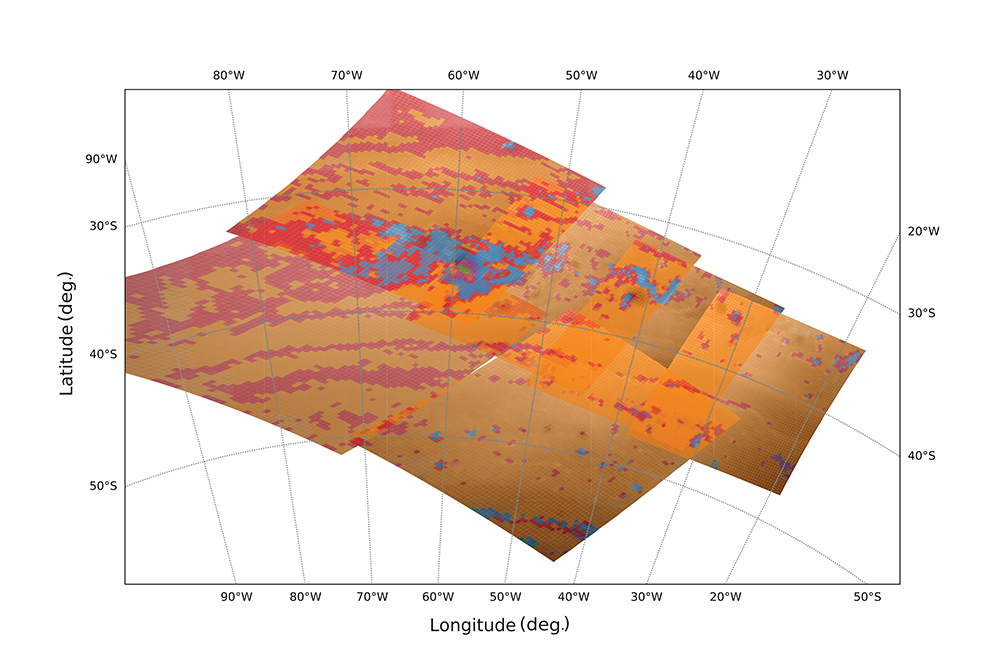}
\caption{{\bf Cloud distribution as mapped by PlanetNet across six overlapping data sets.} Colours are identical to Figure~\ref{fig:specs}.
It is apparent that the SR feature (blue) occurs in the vicinity of dark storms. \label{fig:stich}}
\end{figure}

\begin{figure}
\centering
\includegraphics[width=\textwidth]{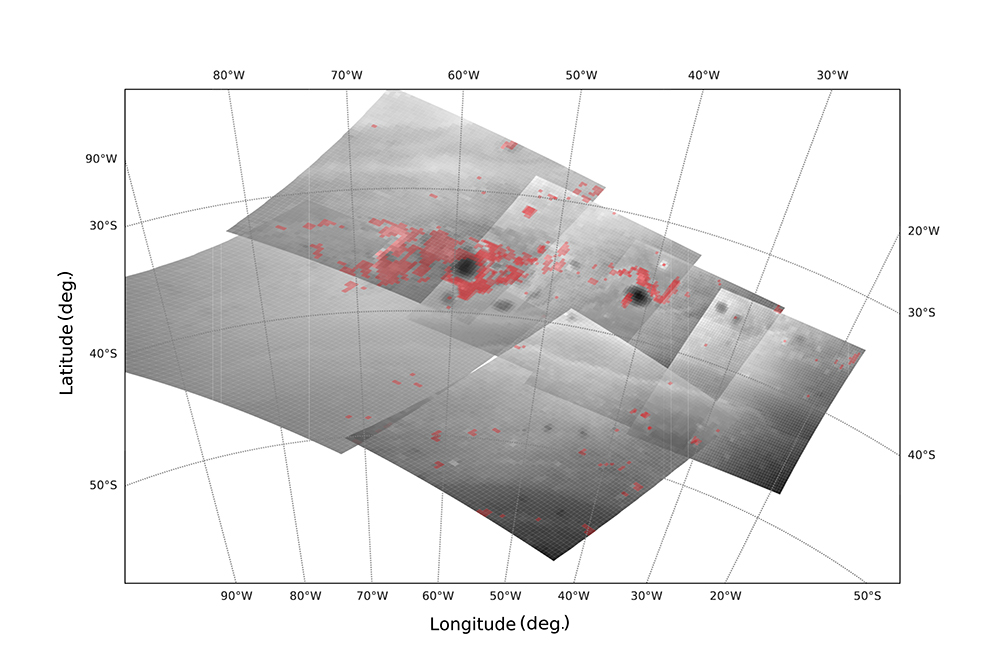}
\caption{{\bf Stormy region as mapped by PlanetNet across six overlapping data sets.} Same as Figure\,\ref{fig:stich} but only showing the 'stormy region' (SR) referred to in the main text and Figure~\ref{fig:specs} \label{fig:stich2}}
\end{figure}
%

\newpage
\paragraph*{Methods}

Here we outline the architecture and training of PlanetNet shown in Figure \ref{fig:chart}. It is a two-stream convolutional neural network (CNN), which analyses the spatial and spectral data of the VIMS data cube separately before combining information in a common, fully connected network. 
Convolutional Neural Networks, also known as translation invariant networks, have been designed specifically for image analysis. As opposed to fully connected deep belief architectures, they are are translation invariant, mean that image features can equally be recognised no matter where they occur in the image. Each CNN layer contains three separate stages: 1) Convolution, 2) non-linear transform (ReLU) and 3) down-sampling (pooling). In the convolution stage, the 2D image is convolved with a set of `filter functions' or kernels. These filter functions are continuously learned from the data during training and provide the decomposition into spatial features (e.g. edges, blobs, etc.). We now convolve each filter (computing the dot product) with the input image to obtain the activation map for each filter. In the second stage, we apply a non-linearity transformation to the activation maps. It is now standard practice to use rectifier linear units (ReLUs) or leaky ReLU as CNN non-linearity. We here use the classical ReLU which is given as $f(x) = \text{max}(0,x)$ and effectively removes all negative entries of the activation map, increasing the non-linearity of the transform. Finally, we down-sample in size, or `pool', the activation maps. Here we use simple $2 \times 2$ max-pooling, where the maximum value of a $2 \times 2$ grid is retained. This procedure is now repeated for the second CNN layer.

The CNN presented here is classical in all aspects apart from the inclusion of two separate and simultaneous streams that treat the spatial and spectral data independently before combining both streams in a fully connected layer. Due to the intrinsic correlations in the spectral channel, we opt for a 1D CNN architecture instead of fully connected layers only. We refer the interested reader to the standard literature on neural networks and CNNs (e.g. \cite{Bishop:2006:PRM:1162264,Bengio:2009kb,Goodfellow-et-al-2016}). The neural network has been implemented in Tensorflow\cite{tensorflow2015-whitepaper}. 
The input data on which training and subsequent classification is performed is denoted by 

\begin{equation}
{\bm x} \in \mathbb{R}^{{H^{0}\times W^{0} \times S^{0}}}
\end{equation}

where x is a three-tensor of dimensions $H^{0}$, $W^{0}$ and $S^{0}$, denoting the height, width and spectral axes of the data cube. The zero index denotes the input (i.e. data) layer of the neural network. Here and throughout we define tensors and matrices by lower and upper-case bold letters respectively. For ease of notation, we also define the sub-tensors ${\bm x}_\lambda={\bm x}^s$ and ${\bm x}_\phi={\bm x}^{(H\times W)}$ denoting the spectral and spatial information only. 
As shown in Figure \ref{fig:chart}, the CNN features two convolution and pooling layers ($l \in {1,2,...,L}$) for each spatial and spectral channel. These can be defined as follows

\begin{equation}
{\bm y}_i^l= \text{pool}_P \left [ \sum_j^{N^{l-1}} \beta({\bm y}_j^{l-1}\otimes {\bm w}_{i,j}^l+{\bm b}_i^l ) \right ]
\end{equation}

where ${\bm y}_i^l$  is the CNN output at layer l and the feature map i. For the spatial and spectral channels, ${\bm y}_i^l$   constitutes a 2D and 3D tensor (${\bm y}_{\lambda,i}^l \in \mathbb{R}^{S^l\times B^l } $ and ${\bm y}_{\phi,i}^l \in \mathbb{R}^{H^l \times W^l \times B^l }$) respectively with B denoting the individual training batch size. ${\bm y}_j^{l-1}$ is the feature map of the previous layer and ${\bm w}
_{i,j}^l$ the corresponding kernel of size $K^l$, and $\otimes$ designates linear convolution. We define the biases as ${\bm b}_i^l$ and note that ${\bm y}^0={\bm x}$. The activation function, $\beta$, is defined as a linear rectifier (ReLu) unit and a max-pooling has been used for both layers and channels. 

After both convolution layers, the spatial and spectral channels are combined to form the input of a fully connected network. We define the fully connected part of the neural network as follows

\begin{equation}
{\bm z}=\beta\{{\bm W}^c \cdot \beta [{\bm W}^{c-1} \cdot ({\bm y}_\phi^L \odot {\bm y}_\lambda^L )+{\bm b}^{c-1} ] +{\bm b}^c\}
\end{equation}

\noindent where $\odot$ denotes the concatenation of ${\bm y}_\phi^L$ and ${\bm y}_\lambda^L$, the outputs of the final convolution layers. The fully connected layer indices are given by $c \in {1,2,\dots,C}$ and ${\bm W}$ and ${\bm b}$ are the weight matrices and biases respectively. The number of neurons per layer is defined by the hyperparamter ${\bm M}^c$. For simplicity, we refer to the set of free parameters as $\vartheta=\{{\bm w}^l,{\bm b}^l,{\bm W}^c, {\bm b}^c\}$.

Finally, we map ${\bm z}$ to our binary training labels, $\theta$ , using a softmax regression layer. To train, we now minimise the cross-entropy of the system.

\begin{equation}
Cost= \frac{-1}{n} \sum_x \left [ \theta \text{ln}({\bm a})+(1-\theta)  \text{ln}(1-{\bm a})\right ]
\end{equation}

\noindent where $n$ is the total number of training data (or batch size),  $\theta$ the vector of training labels and $a= \varsigma({\bm z})$, where $\varsigma(.)$ is a sigmoid activation function. 



Each Cassini/VIMS data cube consists of two spatial and one spectral axis. The spatial dimensions of the cube are 64$\times$64 pixels and 256 wavelength points. As input to PlanetNet, we require spectra/spatial pairs. These are generated as follows: For every pixel, the full spectrum is extracted, i.e. $S^0=256$. Surrounding the central pixel, we also extract the spatial information. The patch size is defined by $H^0$ and $W^0$ (which we later determined to be 20 pixels each). Here the spatial patch is the mean average over all wavelengths. We extract spatial/spectral pairs for all 4096 pixels. Note that spatial patches are overlapping from adjacent pixels. 

In order to train the neural network, we must obtain labeled data. This can be obtained in two ways: 1) labelling individual pixels by hand to mark predominant features or 2) use a statistical clustering algorithm to obtain estimated labels on which the network can learn and improve. 
Here we use spectral clustering on the data cube initially presented by \cite{baines09} to obtain the NH$_3$ signature. Spectral clustering is a preferable clustering algorithm (compared to for example k-means or nearest neighbours) when the structures of clusters are highly non-convex and simple affinity based discriminators break down. 
For spectral clustering, we follow \cite{Yu:2003:MSC:946247.946658,Ng01onspectral} and use the sklearn\cite{scikit-learn} implementation. We define the symmetric graph Laplacian,

\begin{equation}
 \mathcal{\bm L}_{sym}= {\bm D}^{-1/2}  \mathcal{{\bm L}\bm D}^{-1/2}
\end{equation}

\noindent where $\mathcal{L}={\bm D}-{\bm A}$, ${\bm A}$ is the affinity matrix and ${\bm D}$ is the degree matrix. The elements of the affinity matrix are computed using  

\begin{equation}
a_{i,j}=\text{exp}\left (-\frac{1}{2\sigma^2}d^2 [{\bm x}_{\lambda,i},{\bm x}_{\lambda,j}] \right )
\end{equation}

\noindent where $i,j$ are the spatial indices of the data and $d^2$  is the  L$^2$ norm or Euclidian distance given by 

\begin{equation}
    d_{i,j} = \sqrt{\sum{( {\bm x}_{i} - {\bm x}_{j})^2}}
    \label{equ:d}
\end{equation}

\noindent The degree matrix is given by 

\begin{equation}
{\bm D}=\text{diag} \left ( \sum_j a_{i,j}  \right ).
\end{equation}

\noindent Once the graph Laplacian is calculated, we decompose $\mathcal{L}_{sym}$ into its eigenvalues and eigenvectors and rank them from smallest to largest eigenvalue. We estimate the likely number of clusters present in the data by using the eigenvalue heuristic\cite{Luxburg:2007:TSC:1288822.1288832} 

\begin{equation}
N_{cluster}=\text{max}(\epsilon_{i+1}-\epsilon_i )    ~~~~~  \text{for}~i=[1,2,...,N_\epsilon-1]
\end{equation}

\noindent We now use k-means clustering on the matrix of eigenvectors to divide the data into $N_{cluster}$ labeled partitions. This provides us with training labels for each spatial coordinate $(i,j)$.

Since these cluster labels are estimated, some uncertainty in the cluster edges may persist. To mitigate cluster-edge uncertainties, we prune the edges and discard any pixel that lie along a cluster boundary. Finally, the remaining data-label pairs are split into 70\% training and 30\% validation sets. 



The neural network is trained for $2\times10^4$ iterations with mini-batch sizes of 100 spectral/spatial pairs at a time. We train on the training set only and use the validation set as unseen data to check the network's ability to generalise over new data and to gauge any over-fitting. Over-fitting would generally be observed in systems with decreasing cross-entropy but static or decreasing validation accuracy. To mitigate over-fitting, we employ a 30\% dropout rate across all free variables, $\vartheta$, and a relatively slow learning rate of 0.001. We find that convergence in cross-entropy is usually observed around $1.5\times10^4$ training steps or less with a monotonically decreasing cross-entropy and increasing validation set accuracy, indicating good convergence and no over-fitting. Training takes 60 minutes on a 6-core Intel Xeon E5 (3.5Ghz) CPU or $\sim$5 minutes on a Nvidia Tesla K40 GPU. 

In order to optimise the size of the neural network, we ran a grid of 1000 hyperparameter sets. We define the hyperparameters as: the size of the spatial-patch surrounding the central pixel (red square in Figure \ref{fig:chart}, $H_{0}$ and $W_0$); The number of feature maps for both spatial and spectral channels ($N_\phi^l$ and $N_\lambda^l$), the kernel size ($K_\phi^l$ and $K_\lambda^l$), the down-sampling pool size ($P_\phi$ and $P_\lambda$) and the layer sized of the fully-connected network ($M^c$).
We define a symmetric square of 20$\times$20 pixels ($H^0$,$W^0$) as input and  $N^1=15$ and $N^2=40$ feature maps for both spatial and spectral channels for the first and second layers respectively. Both layers have kernel sizes of $K_\phi^l= 4\times4$ and $K_\lambda^l= 4\times1$ for the spatial and spectral channels respectively.  We use a factor 2 down-sampling with pooling sizes $P_\phi=2\times2$ and $P_\lambda=2\times1$. We summarise the hyperparameter values in Supplementary Table 1.



We investigate the training and prediction accuracies after PlanetNet has been trained on the training cube (TC). Before training, the TC was sub-divided into 70\% training data and 30\% test data. The test data is not shown to the algorithm during training and is used to independently verify the classification accuracy and provide a diagnostic against over-training. In the case of over-training, the classification accuracy of the training set will steadily increase over the training period, whereas the test set accuracy remains unchanged or decreases. In this case, the neural network is memorising the training set rather than learning how to generalise. Supplementary Fig. 11 shows the training (red) and test (purple) accuracies as function of training duration. Both curves are monotonically increasing and show no signs of overfitting. The training and test sets reach $\sim 95\%$ and $\sim 90\%$ classification accuracies. Furthermore, the same figure plots the cross-entropy (loss function) of PlanetNet during training.  

After PlanetNet has been trained, we further verify the classification accuracy by testing the NN on a re-sampled, ``synthetic'' data set. We refer to this data set as re-sampled test cube (RTC). The RTC was generated by re-sampling the training cube in such a way as to preserve the statistical properties of the data but to appear as a new, unseen data to the NN. First, the TC was divided according to the labels obtained by the spectral clustering. Following the previous notation, we refer to the TC data pertaining to a label as ${\bm x}_\theta$, where $\theta$ is the index of the label. Within each label subset, we randomised the order of each element of ${\bm x}_\theta$ to break any close spatial connections in the data. Then we re-sample each train set, both in the spatial ${\bm x}_\theta, \phi$ and spectral  ${\bm x}_\theta,\lambda$ dimensions. 
For the spatial dimension, we transpose ${\bm x}_\theta, \phi$, which is equivalent to a 90 degree rotation. 
This rotation preserves the underlying properties of the data but appears as new, unseen data to the NN. The spectral information of each element in ${\bm x}_\theta,\lambda$ was replaced with the average of the two adjacent spectra in the RTC. Given that all spectral positions in the RTC are randomised, the newly generated spectra will effectively sample from the distribution of the cluster.  

We now let PlanetNet detect cluster labels on RTC using spectral and spatial information, and additionally with spectral or spatial information only. Supplementary Fig. 12 shows the RTC classification for spatial and spectral data. The left shows the `ground-truth' labels, middle the predicted labels by PlanetNet and right is the error matrix of pixels miss-classified. PlanetNet achieves a 93\% classification accuracy, which is in agreement with the training/test data classification accuracies.

We now run the classification for spatial-only and spectra-only cases by setting ${\bm x}_\theta,\lambda = 0$ and ${\bm x}_\theta, \phi = 0$ respectively. This results in very poor cluster label recognition by PlanetNet (48\% and 32\%, Supplementary Figures 13 \& 14, respectively). However we note that this is to be expected for models that are trained on both, spatial and spectral pairs. Suppressing half of the network's information results in erroneous predictions. Hence, if the user wants to use, say, spectral data only for their classification, PlanetNet should be re-trained on spectral-only data.



In addition to our analysis using PlanetNet, we have also conducted a classical principal component analysis\cite{Jolliffe} of the training cube to further verify the classification derived by PlanetNet. In a recent analysis, Griffith et al.\cite{griffith19} has shown that PCA analysis of hyperspectral images of Titan can unveil structural information that otherwise remains hidden in spectral noise. 
PCA is complementary to the main analysis using PlanetNet as principal components probe a different statistics of the data and can hence verify the veracity of PlanetNet's analysis. Whereas the spectral clustering of PlanetNet identifies clusters of similar spatial/spectral regions, PCA decomposes the data along the axes of highest variances. In other words, PlanetNet is sensitive to the statistical distance (equation (\ref{equ:d})) between data points and hence clusters the data according to their `likeness', whereas PCA on the other hand decomposes the data by its variance, broadly speaking its `dissimilarity'. Should distinct spectral classes exist, they will contribute to the variance of the data cube and should hence be separable as individual principal components. Both methods should converge to finding the same regions. We find this to be the case, verifying the validity of our decomposition of the training cube.  
In Supplementary Figure~12 we show the first four principal components of the 0.88 - 1.66\,$\mu$m spectral region alone. These regions probe the albedo of the clouds and are hence most sensitive to the high altitude atmospheric variations. The ``S'' shaped  NH$_3$ ice feature\cite{baines09} is clearly visible in the second to fourth component. Supplementary Figure~13 shows additional PCA maps of the full wavelength range and the thermal emission component of the spectrum ($>$4\,$\mu$m) is shown in Supplementary Figure~14. Both maps clearly show the presence of the SR region in their second principal components. When considering the full wavelength range (Supplementary Figure~13) we see the dark storm to be most prominent in the 1$^{st}$ component, whereas Supplementary Figure~14 shows the SR region prominently both in the first and second components. This suggests that spectral variance in the thermal emission part of the spectrum is strongly driven by the SR region and indicative of turbulent upwelling. Future work will explore the addition of component separation methods to further augment feature detection sensitivities.  




\newpage 
\section*{Supplementary Information}

\subsection*{Euclidean Distance maps}

Additional maps showing individual components as recognised by PlanetNet as well as statistical distance plots. For more information, please refer to the Method section of the main text.

\begin{figure}[h]
\centering
\includegraphics[width=1.0\columnwidth]{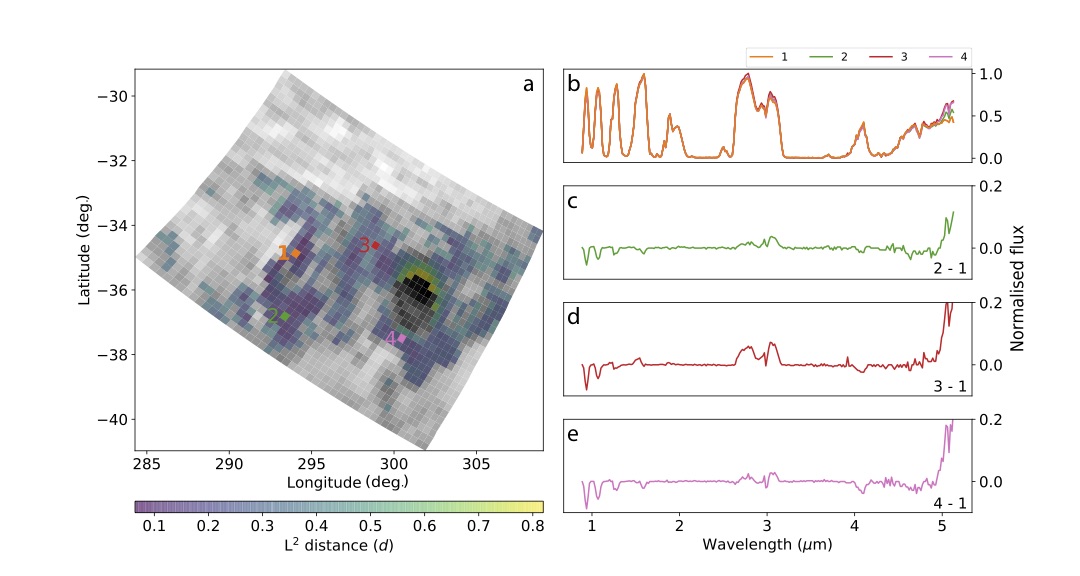}
\caption{{\bf Map of spectral differences inside the storm region (SR)}. A: Map of the SR as function of Euclidian distance (L$^2$ norm) from the peak emission of the ``S'' shaped ammonia ice feature, here labelled 1 (orange). The map shows the statistical similarity of each spectrum compared to (1) with low values being more similar and high values being less similar. B: Spectra corresponding to the pixels labelled on the map. C - E: Spectral differences between spectra (2), (3), (4) and (1) respectively. \label{fig:SR_feature_distance}}
\end{figure}

\begin{figure}[h]
\centering
\includegraphics[width=1.0\columnwidth]{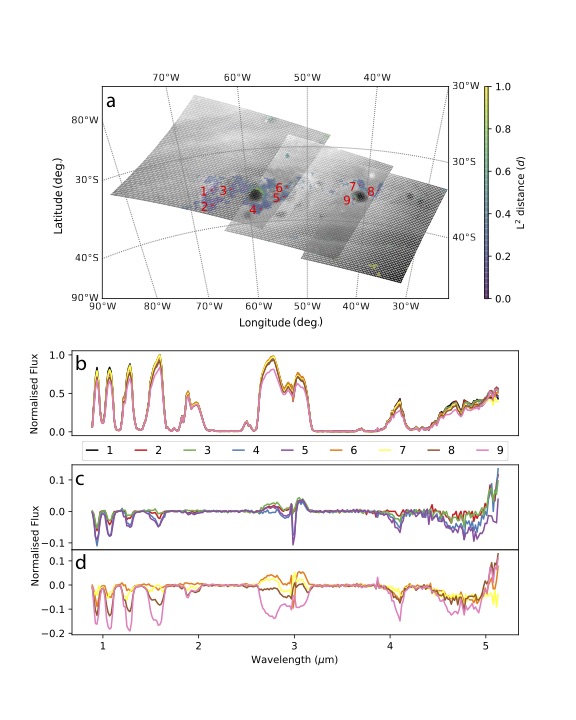}
\caption{{\bf Map of spectral differences inside the storm region (SR) across all six data sets.}
A: Same Euclidean distance map to figure~\ref{fig:SR_feature_distance} but for the larger area shown in figure\,4 of the main text, encompassing the stormy region of the smaller storm at 42$^\circ$W. All Euclidian distances measured are relative to the spectrum marked as (1), with small distances indicating a higher degree of similarity. B: Spectra pertaining to the pixels marked in the map above. C \& D: Spectral differences between spectra in the middle plot and spectrum (1).   \label{fig:SR_stich_spectra}}
\end{figure}

\begin{figure}[h]
\centering
\includegraphics[width=\columnwidth]{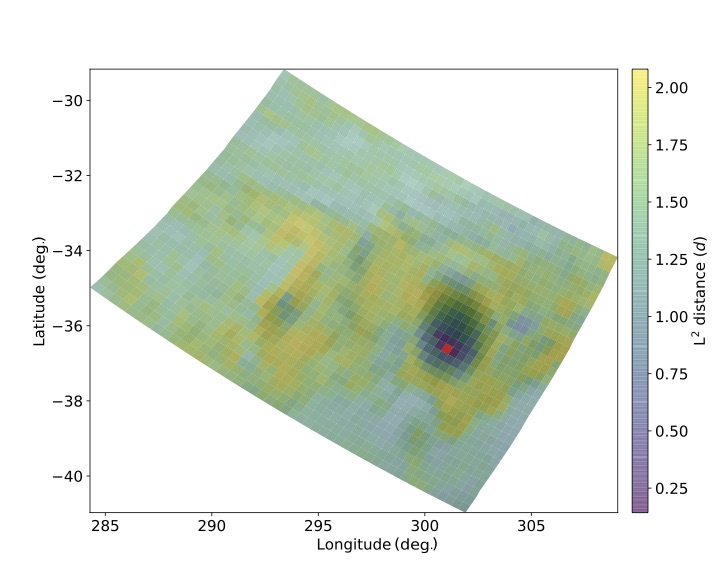}
\caption{{\bf Map of the Euclidian distance (L$^2$ norm) from the centre of the dark storm (red pixel) for all spectra (independent of cluster label).} Similar to supplementary figures\,\ref{fig:SR_feature_distance} \& \ref{fig:SR_stich_spectra} the smaller the distance the higher the similarity between two spectra. The SR region surrounding the dark storm stands out as being statistically distant (i.e. dissimilar) from dark storm signatures. \label{fig:darkstorm_distance}}
\end{figure}

\clearpage
\subsection*{Principal Component Maps}

Maps showing the results of the Principal Component Analysis (PCA) explained in the Method section of the main text. 

\begin{figure}[h]
\centering
\includegraphics[width=\textwidth]{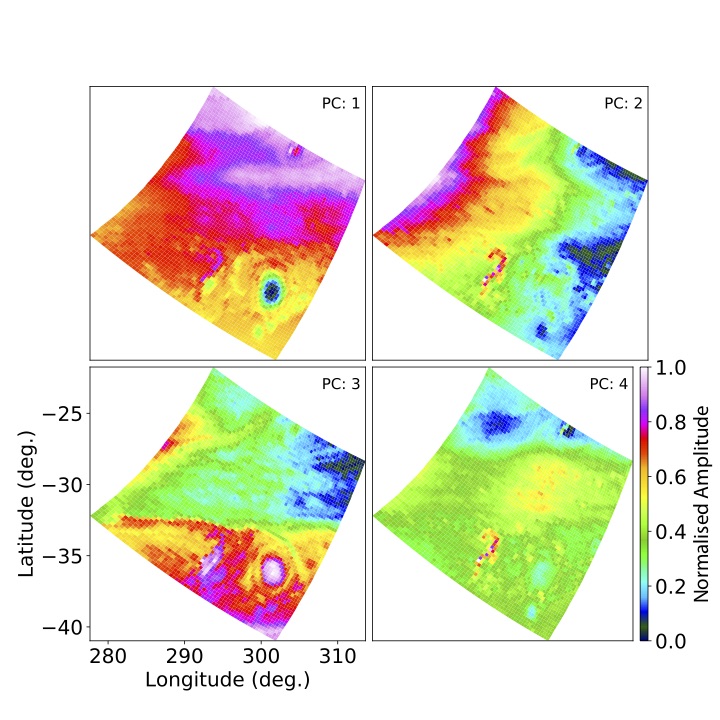}
\caption{{\bf Spatial map of the principal component analysis of the 0.88 - 1.66\,$\mu$m spectral region.} Shown are the first four strongest principal components (PC). PC\,1 is dominated by the dark storm, providing most variance to the spectral map. The 2$^{nd}$ to 4$^{th}$ PCs show the ammonia ice S-shape feature. Each principal component map is normalised to unity for clarity. \label{fig:pca_088-166}}
\end{figure}

\begin{figure}[h]
\centering
\includegraphics[width=\columnwidth]{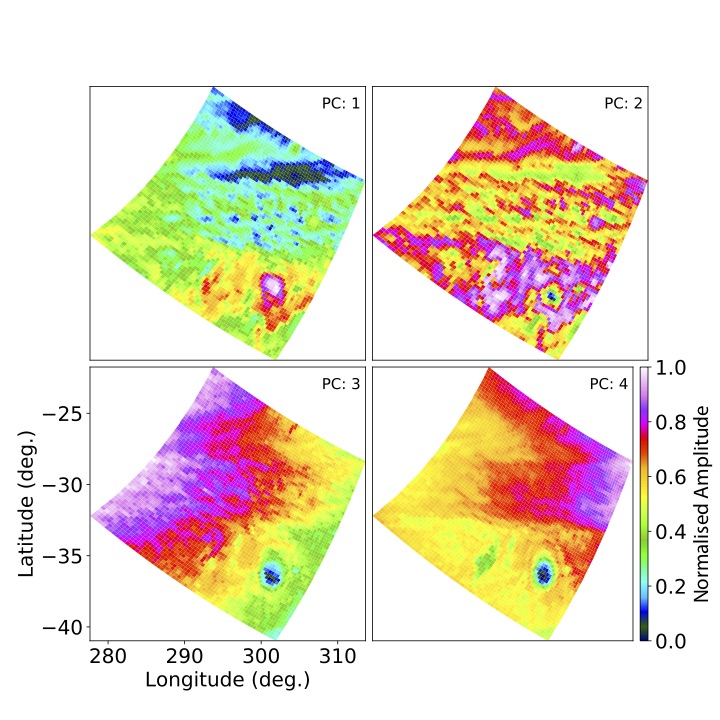}
\caption{{\bf Spatial map of the principal component analysis of the full wavelength range of the instrument.} Clearly visible are the dark storm, the extended SR feature and the high-altitude CH$_4$ clouds. \label{fig:pca_full}}
\end{figure}

\begin{figure}[h]
\centering
\includegraphics[width=\columnwidth]{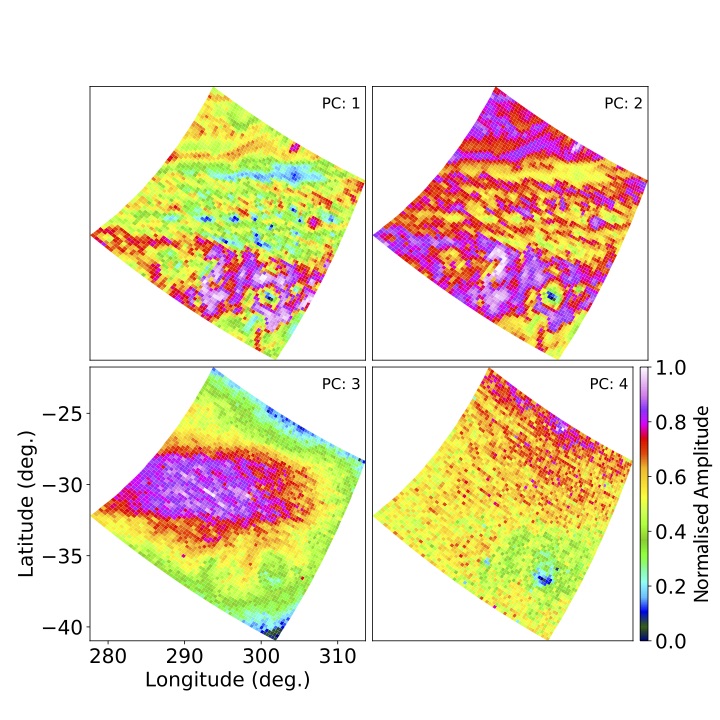}
\caption{{\bf Spatial map of the principal component analysis of the 4.0 - 5.1\,$\mu$m wavelength range}. This range depicts the thermal emission component of the spectrum. PCs 1\,\&\,2 show the storm (SR) region surrounding the dark storm. \label{fig:pca_SR}}
\end{figure}

\clearpage
\subsection*{Additional PlanetNet component maps}

Additional maps showing individual components as recognised by PlanetNet. For more information, please refer to the main text.

\begin{figure}[h]
\centering
\includegraphics[width=\columnwidth]{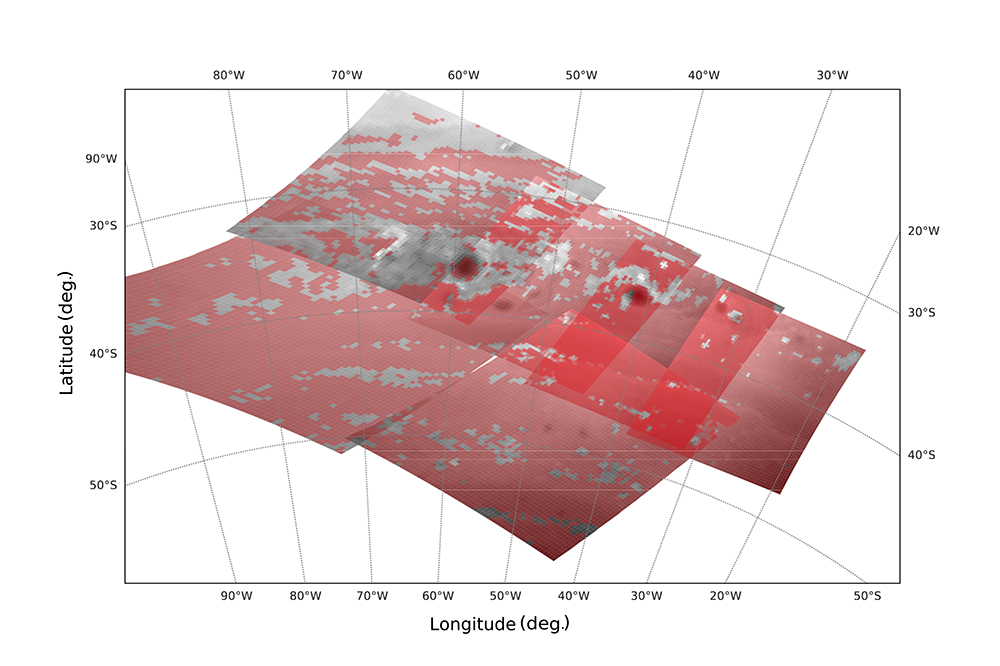}
\caption{{\bf Map of the second spectral/spatial feature identified by PlanetNet}. This feature corresponds to the orange shaded region (spectrum 2) in Figure 2 of the main text.\label{fig:stich3}}
\end{figure}

\begin{figure}[h]
\centering
\includegraphics[width=\columnwidth]{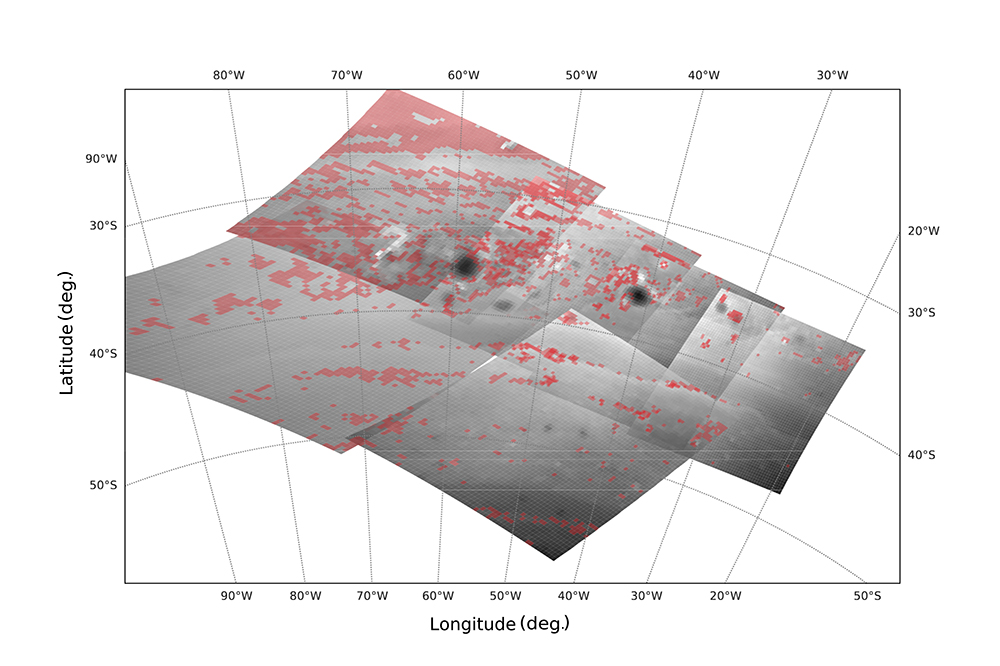}
\caption{{\bf Map of the third spectral/spatial feature identified by PlanetNet}. This feature corresponds to the red shaded region (spectrum 3) in Figure 2 of the main text.\label{fig:stich4}}
\end{figure}

\begin{figure}[h]
\centering
\includegraphics[width=\columnwidth]{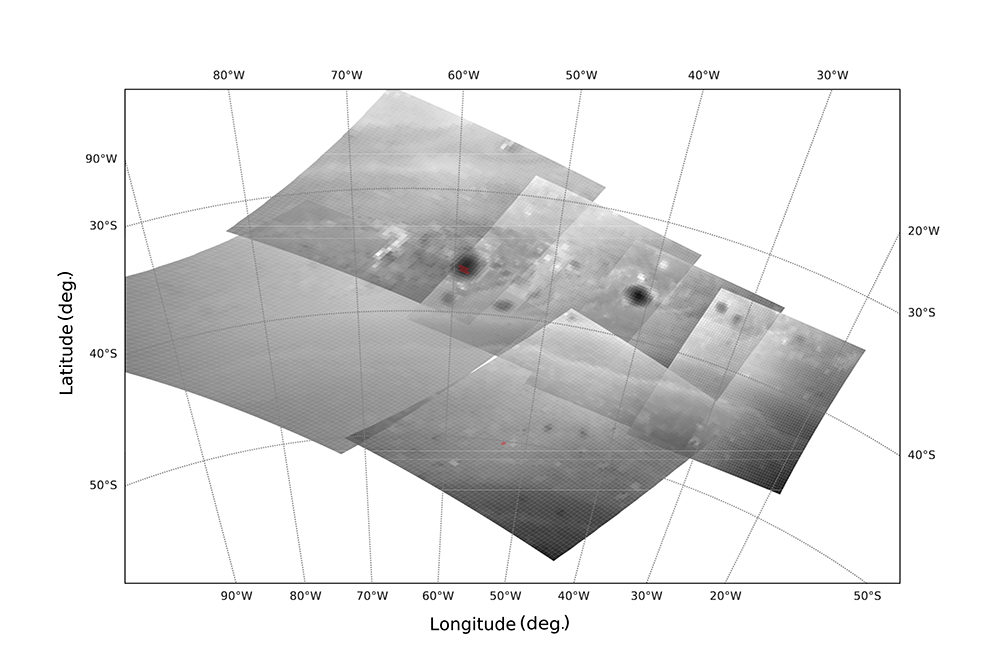}
\caption{{\bf Map of the fourth spectral/spatial feature identified by PlanetNet}. This feature corresponds to the green shaded region (spectrum 5) in Figure 2 of the main text. \label{fig:stich5}}
\end{figure}

\begin{figure}[h]
\centering
\includegraphics[width=\columnwidth]{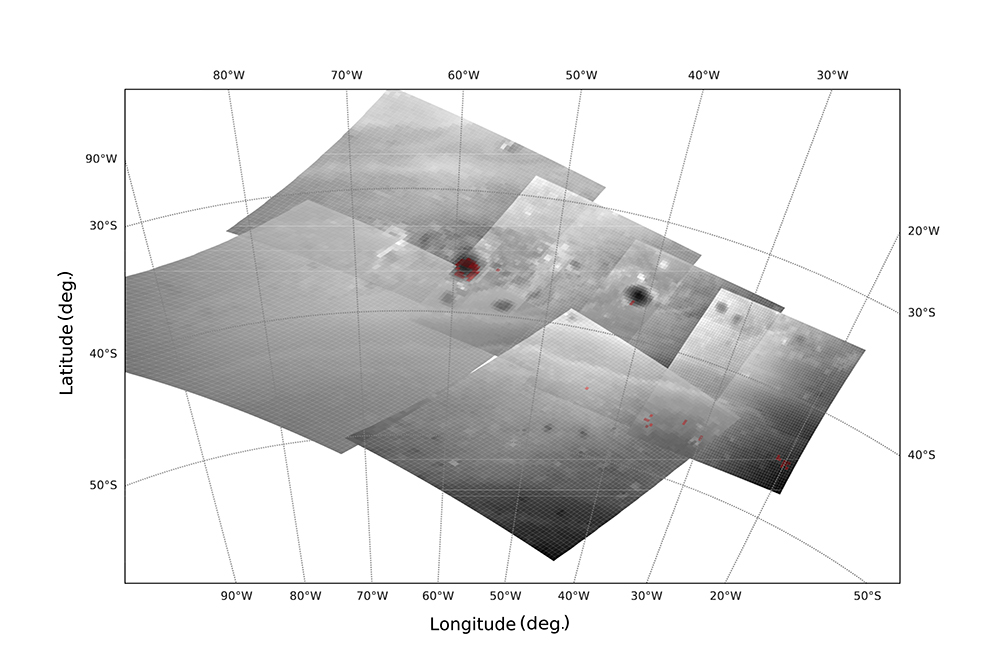}
\caption{{\bf Map of the fifth spectral/spatial feature identified by PlanetNet}. This feature corresponds to the purple shaded region (spectrum 6) in Figure 2 of the main text.\label{fig:stich6}}
\end{figure}

\clearpage
\subsection*{Training and validation}

Additional plots and tables pertaining to the training and validation outlined in the Method section of the main text.

\begin{table}[h]
\center
\caption{{\bf Summary of PlanetNet hyper-parameters and their values.} \label{tbl:hyper}}
\begin{tabular}{l | l}
Hyperparameters &	Value \\\hline
$H^0$, $W^0$	& 20,20 \\
$N^1, N^2$ & 15,40 \\
$K_\lambda^l,K_\phi^l$  & $4\times1$,$4\times4$ \\
$P_\phi,P_\lambda$ & $2\times2$,$2\times1$ \\
$M^1,M^2$ & 1024, 10 \\
\end{tabular}
\end{table}

\begin{figure}[h]
\centering
\includegraphics[width=\columnwidth]{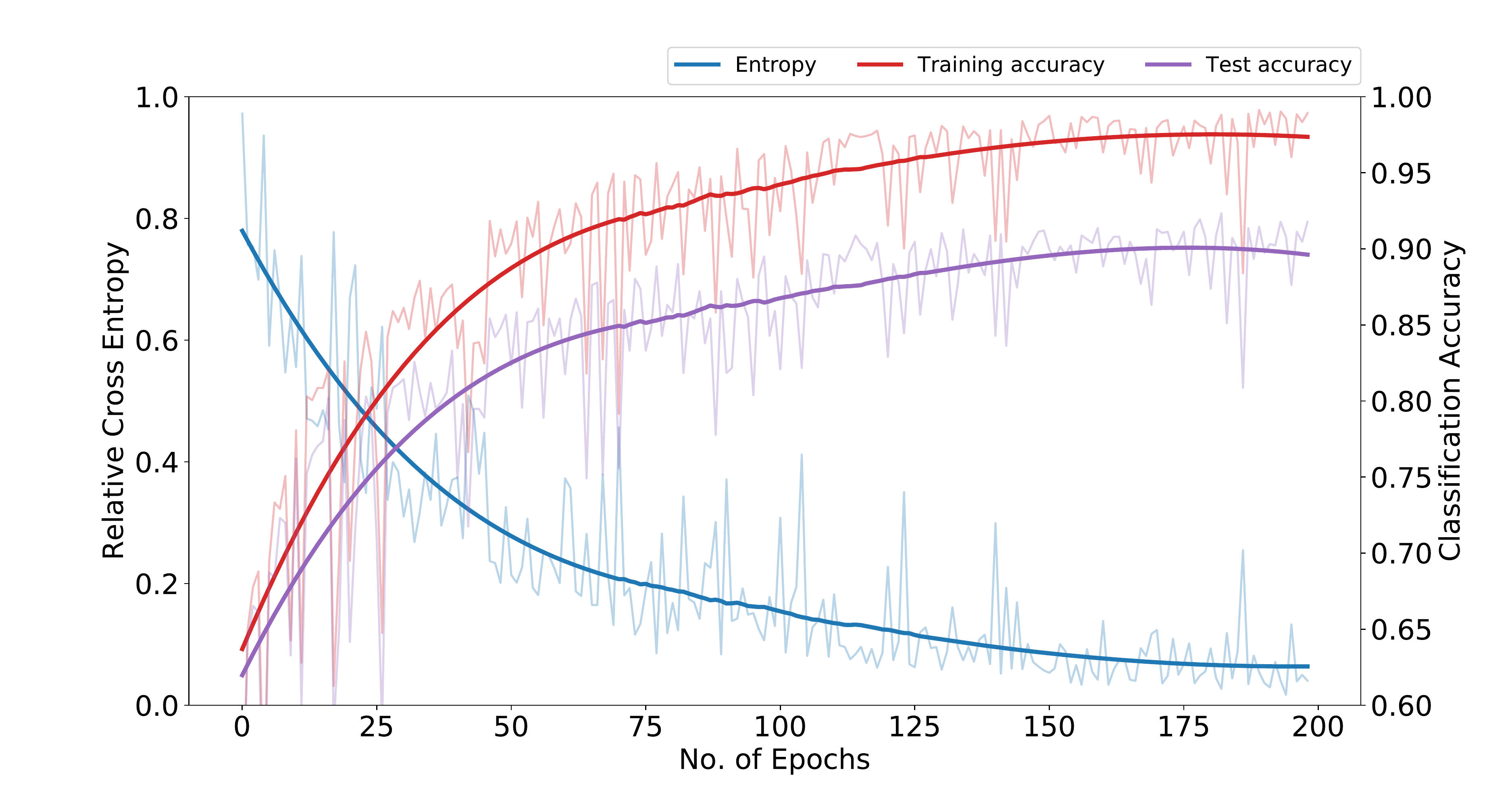}
\caption{{\bf PlanetNet accuracies as function of training duration}. Left Axis/Blue: Relative cross entropy as function of training duration. Right axis: Training (red) and Test (purple) set accuracy as function of training duration. Bold lines present a smoothed average. \label{fig:entropycurve}}
\end{figure}

\begin{figure}[h]
\centering
\includegraphics[width=\columnwidth]{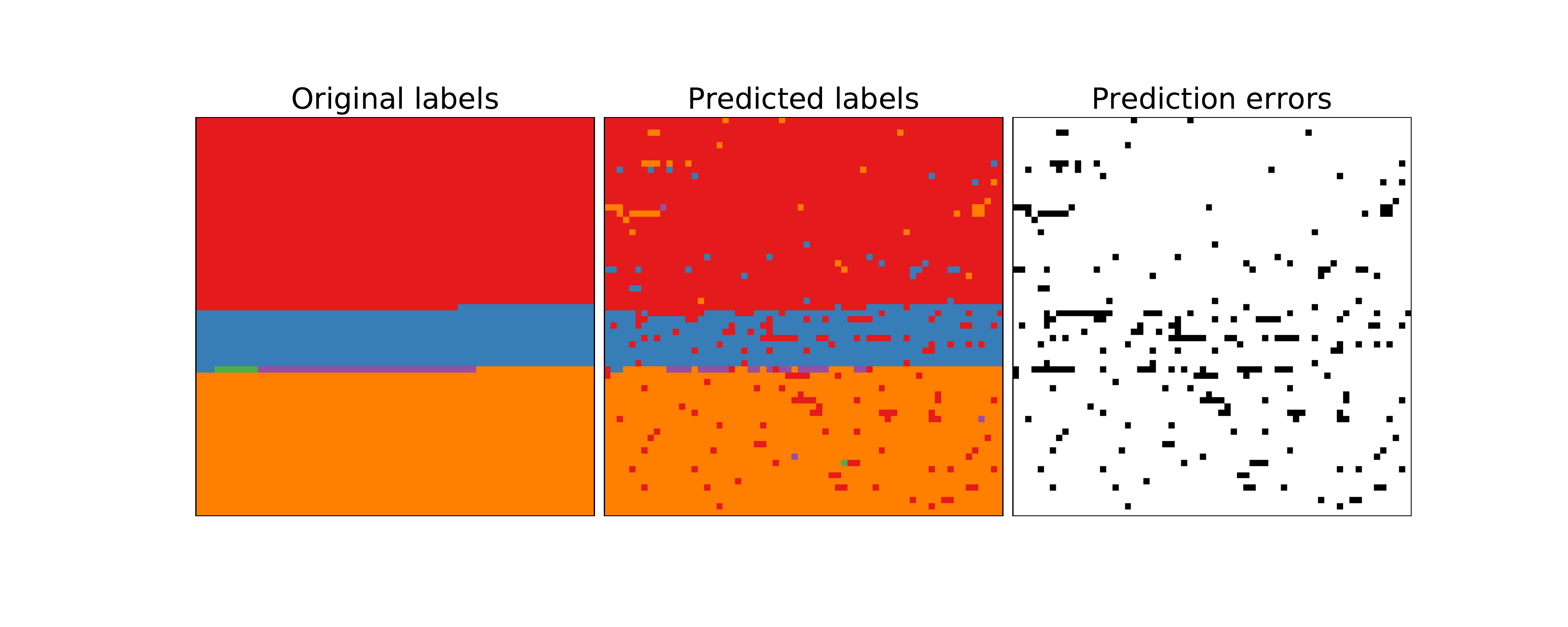}
\caption{{\bf PlanetNet recognition accuracy test on the re-sampled training cube (RTC)}. The RTC is shown in the left panel (Original Labels) with the PlanetNet predicted labels in the centre and the error-matrix (Prediction errors) showing the wrongly identified pixels.  94\% of all pixels were correctly identified.  \label{fig:resampled}}
\end{figure}

\begin{figure}[h]
\centering
\includegraphics[width=\columnwidth]{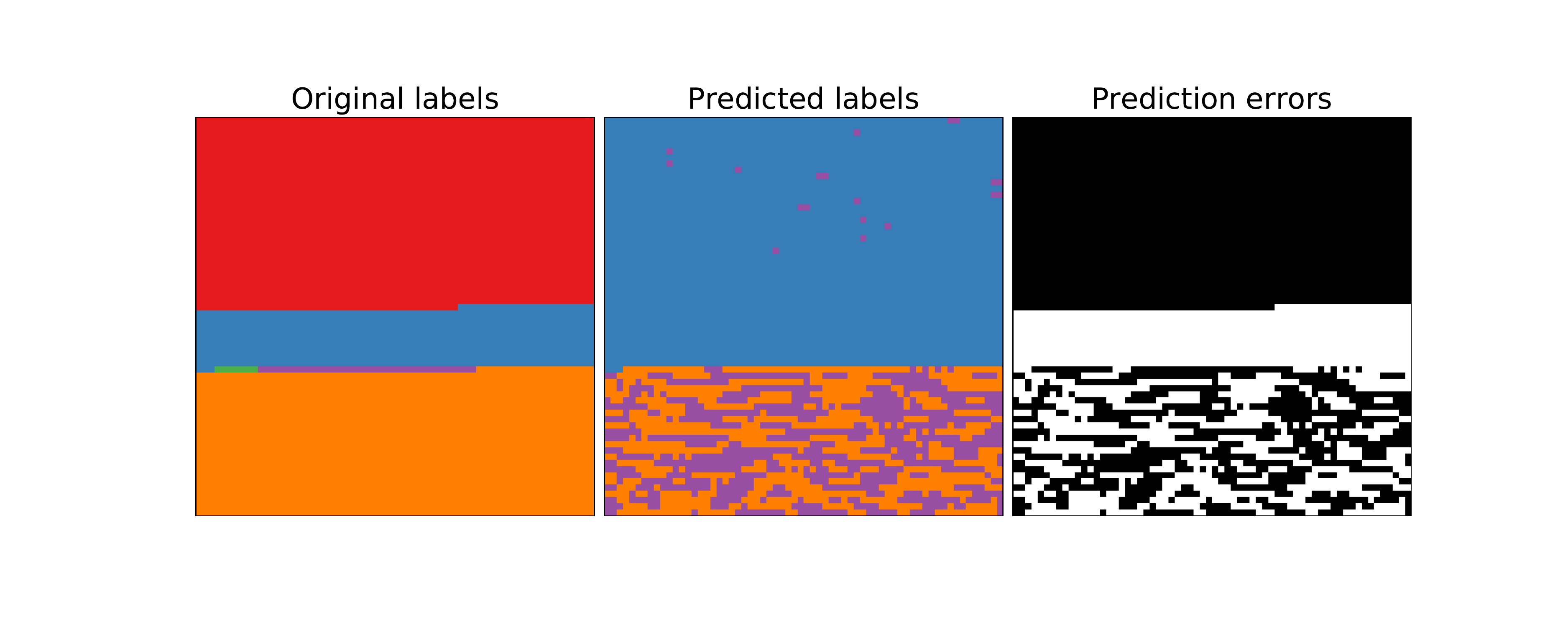}
\caption{{\bf PlanetNet recognition accuracy test on the re-sampled training cube (RTC) when only the spectral information was considered}. The layout is identical to supplementary figure\,\ref{fig:resampled} but here the spatial information was artificially set to zero, resulting in a very low recognition accuracy of 32\%. \label{fig:resample_speconly} }
\end{figure}

\begin{figure}[h]
\centering
\includegraphics[width=\columnwidth]{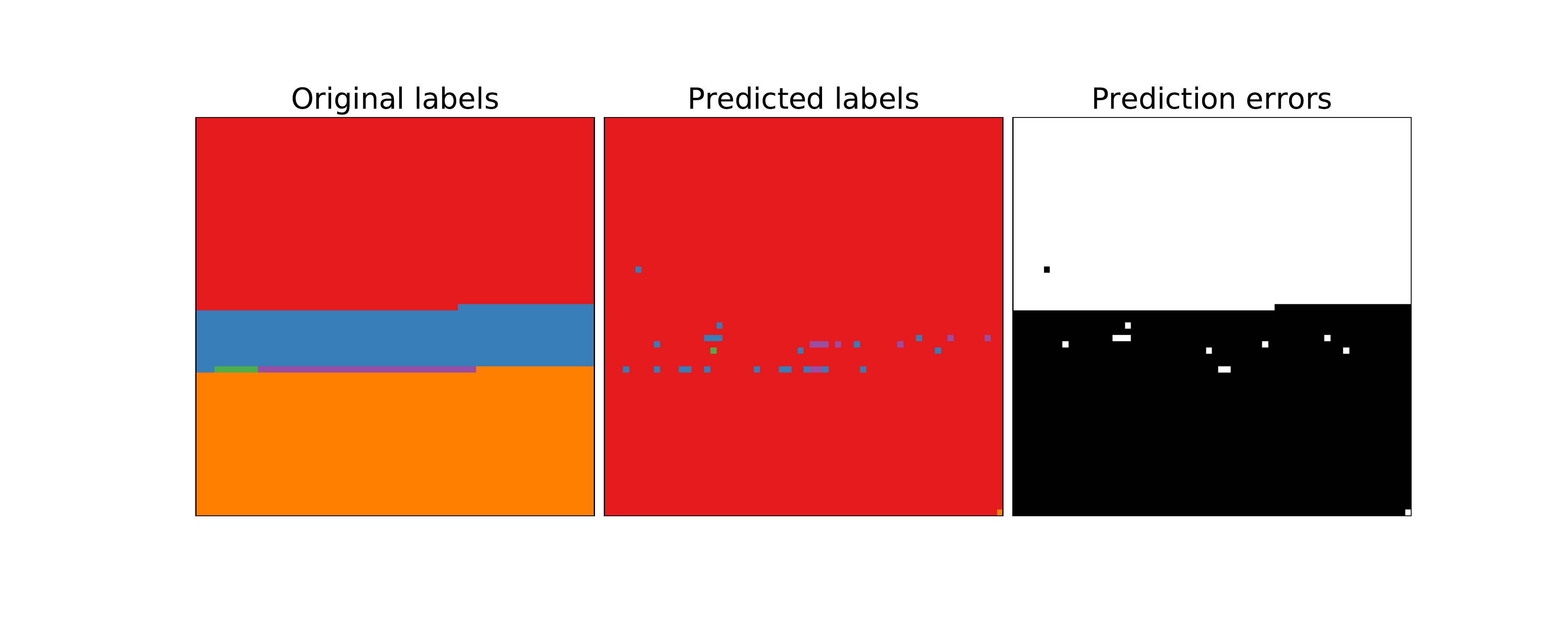}
\caption{{\bf PlanetNet recognition accuracy test on the re-sampled training cube (RTC) when only the spatial information was considered}. The layout is identical to supplementary figure\,\ref{fig:resampled} but here the spectral information was artificially set to zero, resulting in a very low recognition accuracy of 48\%. \label{fig:resample_maponly} }
\end{figure}

\end{document}